\documentclass{iopconfser}

\usepackage[usenames,dvipsnames]{xcolor}
\usepackage{amsmath,amssymb,epsf}
\usepackage{amsfonts,amsbsy,amscd,stmaryrd, wasysym}
  \SetSymbolFont{stmry}{bold}{U}{stmry}{m}{n}
  \usepackage{mathtools}
  \usepackage{subcaption}
\usepackage{latexsym,amsopn}
\usepackage{amsthm}
\usepackage{esint}
\usepackage{mathrsfs}
\usepackage{slashed} 
\usepackage{enumitem}
\usepackage{orcidlink}
\allowdisplaybreaks

\usepackage{soul, todonotes}

\usepackage[utf8]{inputenc}
\usepackage[T1]{fontenc}   

\usepackage{graphicx}
\usepackage{multirow}
\usepackage{float}

\newcommand{\ben}{\begin{arabicenumerate}}  
\newcommand{\een}{\end{arabicenumerate}}

\newtheorem{theorem}{Theorem}[section]

\newtheorem{proposition}[theorem]{Proposition}
\newtheorem{lemma}[theorem]{Lemma}

\theoremstyle{definition}
\newtheorem{definition}[theorem]{Definition}
\newtheorem{remark}[theorem]{Remark}
\newtheorem{example}[theorem]{Example}

\newcommand{\beq}{\begin{equation}}
\newcommand{\eeq}{\end{equation}}
\newcommand{\bea}{\begin{aligned}}
\newcommand{\eea}{\end{aligned}}
\newcommand{\bear}{\begin{array}{rl}}
\newcommand{\eear}{\end{array}}
\newcommand{\bex}{\begin{example}}
\newcommand{\eex}{\end{example}}
\def\bel{\begin{lemma}}
\def\eel{\end{lemma}}
\def\bet{\begin{theoreme}}
\def\eet{\end{theoreme}}
\def\bed{\begin{definition}}
\def\eed{\end{definition}}
\def\ber{\begin{remark}}
\def\eer{\end{remark}}
\def\bep{\begin{proposition}}
\def\eep{\end{proposition}}

\let\origmaketitle\maketitle
\def\maketitle{
  \begingroup
  \def\uppercasenonmath##1{} 
  \let\MakeUppercase\relax 
	\origmaketitle
  \endgroup
}

\def\bar{\overline}

\usepackage{calrsfs}
\DeclareMathAlphabet{\pazocal}{OMS}{zplm}{m}{n}

\def\cB{{\pazocal B}}
\def\cC{{\pazocal C}}

\def\cN{{\pazocal N}}

\def\cV{{\pazocal V}}

\def\cc{{\mathbb C}}

\def\14{\frac{1}{4}}

\def\12{\frac{1}{2}}

\DeclareSymbolFont{boldoperators}{OT1}{cmr}{bx}{n}
\SetSymbolFont{boldoperators}{bold}{OT1}{cmr}{bx}{n}

\makeatletter
\newcommand*{\defeq}{\mathrel{\rlap{%
                     \raisebox{0.34ex}{$\m@th\cdot$}}%
                     \raisebox{-0.4ex}{$\m@th\cdot$}}%
                     =}
                     
\makeatother
\makeatletter
\newcommand*{\eqdef}{=\mathrel{\rlap{%
                     \raisebox{0.34ex}{$\m@th\cdot$}}%
                     \raisebox{-0.4ex}{$\m@th\cdot$}}%
                     }
\makeatother

\def\2Sol{{\rm Sol}_{{\rm L}^{2}}}

\newcommand{\abs}[1]{{\left\vert #1 \right\vert}}

\numberwithin{equation}{section}

\begin{document}
\title{The Teukolsky scalar as a gateway for quantizing gravity on rotating black holes}

\author{Christiane K. M. Klein\,\orcidlink{0000-0003-2398-7226}$^{1}$}
\affil{$^1$University of York, Department of Mathematics, Ian Wand Building, Deramore Lane, York YO10 5GH, United Kingdom}
\email{christiane.klein@york.ac.uk}

\begin{abstract}
The quantization of linearized gravity on black hole spacetimes and the construction of states for that theory is a sought-after, yet difficult achievement. One of the main reasons is the difficulty of reconciling the positivity and gauge invariance of potential states. On Kerr spacetimes, the spin-2 Teukolsky scalars express the same degrees of freedom as the metric perturbations, up to pure gauge and radiative solutions, while avoiding the issue of gauge. In this work, we therefore explore the quantization of Teukolsky scalars on Kerr. We demonstrate that the theory can be written in terms of a formally hermitian Green-hyperbolic operator, allowing the construction of the algebra of observables. The reconstruction scheme for the metric perturbation allows us to keep track of the physical subalgebra. We indicate how this can be used to construct the Unruh state for this theory on any subextreme Kerr spacetime. This is based on a joint work with Dietrich Häfner \cite{HaefnerKlein} in preparation.
\end{abstract}

\section{Introduction}
Recent years have seen dramatic developments in astrophysics, and a major part of these developments is due to the fact that it is now possible to observe black holes, either in the form of their shadows \cite{EHT} or in the form of the gravitational waves emitted by the merger of black-hole binaries or black-hole-neutron-star systems \cite{LIGO}. With an ever-growing catalogue of mergers, and increasing precision of the shadow images, it becomes feasible to employ these observations for tests of the theory of general relativity (GR) and its potential extensions. However, to predict possible signatures of new physics, it is essential to have a good theoretical understanding of the spacetimes in question. This includes in particular rotating black hole spacetimes, since the black holes in our universe are observed to have significant rotation.

But even over half a century after the first derivation of the Kerr metric \cite{Kerr} describing rotating black holes, there is still a number of theoretical questions that remain unanswered. Some of these questions can be addresses within GR. Others are intimately linked to quantum physics, for example the question how the evaporation of a rotating black hole can be described in a self-consistent manner, which connects to other issues such as the information loss paradox \cite{UnruhWald}.

 In the absence of a full theory of quantum gravity, the semi-classical approach can serve as a first step towards comprehending black-hole evaporation and similar phenomena. In this approach, quantum fields are considered on a classical background. The background and the quantum field are coupled by considering the expectation value of the (renormalized) stress-energy tensor of the quantum field as a source term for the Einstein equations of the background. Unfortunately, a self-consistent solution of the semi-classical Einstein equations is still beyond our grasp outside of highly symmetric setups \cite{Sanders,Meda,Gottschalk}. However, progress can still be made outside of these situations by studying quantum fields on a fixed rotating black hole spacetime.

In light of this motivation, our goal is to study the quantization of linearized gravity on rotating black hole spacetimes, following the algebraic approach to quantum field theory. A major difficulty in this endeavour lies in the construction of states. While there are different approaches to the state construction in linearized gravity covering various classes of spacetimes \cite{FH, GWgrav, Ggrav, BDM, CMS}, the construction of states on Kerr spacetimes has not been achieved yet.

\section{The Teukolsky scalars}

One way to circumvent the difficulties arising from the gauge freedom of linearized gravity in Kerr spacetimes is to express the theory in terms of the Teukolsky scalars instead of the metric perturbations.

The Teukolsky scalars arise as contractions of the Weyl tensor with a geometrically adapted null tetrad, or the Weyl spinor with a geometrically adapted spinor dyad. The Bianchi identities for the Weyl tensor in linearized form can be rewritten as a system of equations for the Teukolsky scalars.  In fact, the highest- and lowest-spin components of this system can be decoupled and give the Teukolsky equations of spin $\pm 2$ \cite{Teukolsky1, Teukolsky2}. In the case of vacuum perturbations, one of these components is sufficient to reconstruct the metric perturbations up to pure gauge contributions and certain radiative terms \cite{ Wald, Chrzanowski, PSW, Ori, Pound:2021, TZSHPG, BGL}. 
Moreover, the corresponding Teukolsky operators $T_{\pm 2}$ governing the decoupled equations are normally hyperbolic \cite{Bini}. Hence, they possess unique retarded and advanced Green operators $E_{\pm 2}^\pm$ on any globally hyperbolic extension of the Kerr spacetime \cite{BGP}. Consequently, one obtains a phase space with a (charged) symplectic form $\sigma$, which can serve as the basis for quantizing the theory \cite{DG}.

In spite of this, it is not trivial to quantize the Teukolsky scalars. The reason is that the geometrically adapted null tetrad is not uniquely fixed, but can be rescaled and rotated. If one does not want to fix the scaling in some arbitrary way, this implies that the Teukolsky scalars will transform as $\phi_{\pm 2}\to  z^{\pm 2}\phi_{\pm 2}$ when the tetrad is rescaled by $\abs{z}$ and rotated by $\arg{z}$. Since the tetrad cannot be chosen globally, the Teukolsky scalars thus live in non-trivial associate $\cc\setminus\{0\}$-bundles $\cB(\pm2)$, with the representation of $\cc\setminus\{0\}$ (as a multiplication group) given by $z\to z^{\pm2}$, see \cite{HaefnerKlein, Millet1} for a more detailed account. What is more, these bundles lack a hermitian fibre metric and a complex conjugation $\cC: \cB(\pm 2)\to \cB(\pm 2)$. These structures are usually assumed to exist and play an important role in the quantization of theories described by Green-hyperbolic operators such as $T_{\pm 2}$. 

A way to mitigate this is to consider the operator $T_{2}\oplus\overline{T_{-2}}$ acting on sections of the enlarged bundle $\cV_2=\cB(2)\oplus \overline{\cB(-2)}$ instead. Since $\overline{\cB(-2)}$ is the anti-dual bundle of $\cB(2)$, and $\overline{T_{-2}}$ the anti-dual operator to $T_2$, one can show that this extended bundle possesses a natural hermitian fibre metric, and that the enlarged operator is formally hermitian with respect to it. The enlarged theory can thus be quantized following the well-established procedure for the quantization of formally hermitian Green-hyperbolic operators \cite{ DG,GW, F}: The algebra of observables can be given by the CCR-algebra of the theory, that is as the free unital $*$-algebra generated by (anti-)linearly smeared fields $\Phi(\phi)$ and $\bar{\Phi}(\phi)$, where $\phi\in \Gamma^\infty(\cV_2)$ is a space-compact solution to the enlarged Teukolsky equation, and the canonical commutation relation is given by $[\Phi(\phi), \bar{\Phi}(\psi)]=i \sigma(\phi,\psi) 1\!\!1$.

There are two issues with the quantized theory. The first one is that the natural hermitian fibre bundle on $\cV_2$, while non-degenerate, is not positive. The second is that we artificially doubled the degrees of freedom. We address both issues using the Teukolsky-Starobinsky identities, which relate solutions $\phi_2$ and $\phi_{-2}$ of the Teukolsky equations $T_2\phi_2=T_{-2}\phi_{-2}=0$ if they belong to the same metric perturbations. In addition, we make use of the metric reconstruction scheme for vacuum perturbations, which parametrizes the metric perturbations in terms of derivatives of a Hertz potential $\psi$ that itself satisfies a Teukolsky equation \cite{ Chrzanowski, PSW, Ori, Pound:2021, TZSHPG, BGL}. 
In this way, a section $\phi=(\phi_2, \overline{\phi_{-2}})\in \Gamma^\infty(\cV_2)$ solving $(T_2\oplus \overline{T_{-2}})\phi=0$ and corresponding to a single metric vacuum perturbation can be derived from a Hertz potential $\psi\in \Gamma^\infty(\cB(2))$ satisfying $T_{-2}\psi=0$ by applying differential operators.

We have thus parametrized a physical phase space, allowing us to identify a physical subalgebra within the algebra of observables for the enlarged theory. It remains to construct a state thereon.

\section{A state for the Teukolsky scalars}
In the algebraic approach to quantum field theory, a state is a positive, linear, normalized map from the algebra of observables into the complex numbers.
The class of states that are most straightforward to construct are quasi-free states. A state is said to be quasi-free if it is completely specified by its two-point function. Hence, to construct a quasi-free state, it is sufficient to give a two-point function, often in the form of a bi-distribution, and show that it satisfies a number of properties, namely being a weak bi-solution to the equations of motion, being positive, and agreeing with the symplectic form of the classical phase space up to a symmetric part.

The Unruh state is an example of a quasi-free state. It first appeared in the context of the exploration of the evaporation of Schwarzschild black holes \cite{Unruh}. It is constructed such that at past null infinity, it resembles the Minkowski vacuum, while at future null infinity, it contains the thermal radiation expected from black-hole evaporation. In that sense, it is physically motivated and captures the late-time behaviour expected for the gravitational collapse to a black hole \cite{Hafner}. In addition to that, the Unruh state is well-defined \cite{DimockKay} and Hadamard \cite{DMP}, a property that is widely accepted to single out the class of physically reasonable states. Analogues of the Unruh states have been constructed for real scalar fields on a variety of black hole spacetimes \cite{BrJo, HWZ, Klein}, notably not including Kerr black holes. However, an Unruh state for massless free fermions on Kerr has been constructed in \cite{GHW}.

In all these examples, the Unruh state is defined from its properties on the past (conformal) boundary of the spacetime. This is done by including the algebra of observables of the spacetime injectively into an algebra associated with the (conformal) boundary. A state constructed on the boundary algebra can then be pulled back to the spacetime algebra by the dual map of the injection. The keys to the injective inclusion into the boundary algebra are either scattering maps or decay estimates for the classical theory, which allow to show that there is an injective symplectomorphism from the classical phase space to some symplectic function space on the boundary. 

In the case of the Kerr spacetime, the past boundary consists of past null infinity and the past event horizon of the black hole. Decay results for the Teukolsky scalar are provided by \cite{Millet2}. Together with some careful consideration of the choice of tetrad, they allow the embedding of the bulk algebra into an algebra associated with the boundary. Additionally, we can identify the subalgebra of the boundary algebra into which the physical subalgebra of the spacetime is mapped.

As for other examples of Unruh states, the two-point functions on the past horizon and at past null infinity are then chosen such that they single out positive-frequency modes with respect to the affine parameter of the null geodesics generating the corresponding piece of the boundary \cite{KW}. However, in comparison to previous results, the non-trivial structure of the bundle $\cB(2)$ demands some care. 

In particular, we define the Unruh state only on the physical subalgebra. As a consequence, if  $(\phi_2, \overline{\phi_{-2}})$ is a section of $\cV(2)$ over the boundary which labels a generator of the physical boundary-subalgebra, then $\phi_2$ and $\overline{\phi_{-2}}$ are related to each other by differential operators. Making use of this relation, one can effectively redistribute derivatives to not only ensure the well-definedness of the state, but also its positivity.

The condition on the antisymmetric part of the two-point function can be traded for a second positivity condition, which can be proven in the same way as the first positivity condition. As a result, the carefully crafted two-point function pulls back to the two-point function of a well-defined state on the  physical spacetime algebra of the Teukolsky scalars.

It remains to show the Hadamard property of the Unruh state, i.e. to show that the wavefront set of the two-point function is contained in $\cN^+\times\cN^-$, where $\cN^\pm $ is the future/past lightcone. The proof relies on the Propapgation of Singularities theorem \cite{DH}. It combines the decay estimates \cite{Millet2} and direct computations with a detailed analysis of the null geodesics \cite{HaK}, the analyticity properties of the two-point function first recognized in \cite{DMP}, and the proof for the Hadamard property of passive states \cite{SV}.

Summarizing, we have shown
\begin{theorem}
    The theory of a spin-2 Teukolsky scalar on any subextreme Kerr spacetime can be quantized by using an enlarged classical phase space, and a physical subalgebra can be identified.

    On the physical subalgebra, one can define an analogue of the Unruh state. The state is well-defined and Hadamard on any subextreme Kerr spacetime.
\end{theorem}

This result is an important first step towards a quantization of linearized gravity on rotating black holes.

\section*{Acknowledgements}  The author is funded by the Deutsche Forschungsgemeinschaft (DFG, German
Research Foundation) – Projektnummer 531357976. It is a pleasure to thank Dietrich Häfner and Gerrit Anders for valuable feedback on the draft of this paper.

\bibliographystyle{iopart-num}
\bibliography{teukolsky}
\end{document}